# Molecular responses of mouse macrophages to copper and copper oxide nanoparticles inferred from proteomic analyses

Sarah Triboulet [1, 2, 3], Catherine Aude-Garcia [1, 2, 3], Marie Carrière [4], Hélène Diemer [5], Fabienne Proamer [6], Aurélie Habert [7], Mireille Chevallet [1, 2, 3], Véronique Collin-Faure [1, 2, 3], Jean-Marc Strub [5], Daniel Hanau [6], Alain Van Dorsselaer [5], Nathalie Herlin-Boime [7], Thierry Rabilloud [1, 2, 3]*

1: Pro-MD team, Laboratoire de Chimie et Biologie des Métaux, UMR CNRS-CEA-UJF, Université Joseph Fourier, Grenoble

2: Pro-MD team, CEA Grenoble, iRTSV/LCBM, Laboratoire de Chimie et Biologie des Métaux, UMR CNRS-CEA-UJF, Grenoble

3: Pro-MD team, UMR CNRS 5249, Laboratoire de Chimie et Biologie des Métaux, UMR CNRS-CEA-UJF, Grenoble

4: UMR E3 CEA-Université Joseph Fourier, Service de Chimie Inorganique et Biologique, Laboratoire Lésions des Acides Nucléiques (LAN), Grenoble

5: Laboratoire de Spectrométrie de Masse BioOrganique (LSMBO), Université de Strasbourg, IPHC, 25 rue Becquerel 67087 Strasbourg, France. CNRS, UMR7178, 67037 Strasbourg, France.

6: UMR_S725, INSERM-UdS, EFS-Alsace, 10, rue Spielmann, 67065 Strasbourg

7: IRAMIS, SPAM-LFP (CEA CNRS URA 2453), Saclay

*: to whom correspondence should be addressed:

Laboratoire de Chimie et Biologie des Métaux, UMR CNRS-CEA-UJF 5249, iRTSV/LCBM, CEA Grenoble, 17 rue des martyrs, F-38054 Grenoble Cedex 9, France
thierry.rabilloud@cea.fr

Running title: macrophage responses to copper oxide nanoparticles

## ABBREVIATIONS

BHP: t-butyl hydroperoxide; CHAPS:3-[3- Cholamidopropyl]-dimethylammoniopropane-1-sulfonate; EDTA: ethylenediamine tetraacetic acid; LPS: lipopolysaccharide; MS: mass spectrrometry; MS/MS: tandem mass spectrometry; PBS: phosphate buffered saline: PVP: polyvinyl pyrrolidone; SEM: scanning electron microscopy; TCEP: tris(carboixyethyl)phospine; TEM: transmission electron microscopy

## ABSTRACT

The molecular responses of macrophages to copper-based nanoparticles has been investigated by a combination of proteomic and biochemical approaches, using the RAW264.7 cell line as a model. Both metallic copper and copper oxide nanoparticles have been tested, with copper ion and zirconium oxide nanoparticles as controls. Proteomic analysis highlighted changes in proteins implicated in oxidative stress responses (superoxide dismutases and peroxiredoxins), in glutathione biosynthesis, in the actomyosin cytoskeleton and in mitochondrial proteins (especially oxidative phosphorylation complex subunits). Validation studies by functional analyses showed that the increase in glutathione biosynthesis and in mitochondrial complexes observed in the proteomic screen was critical to cell survival upon stress with copper-based nanoparticles; pharmacological inhibition of these two pathways enhanced cell vulnerability to copper-based nanoparticles but not to copper ion. Furthermore, functional analyses using primary macrophages derived from bone marrow showed a decrease in reduced glutathione levels, a decrease in the mitochondrial transmembrane potential, as well as an inhibition of phagocytosis and of lipopolysaccharide-induced nitric oxide production. However, only a fraction of these effects could be obtained with copper ions. In conclusion, this study showed that macrophage functions are significanlty altered by copper-based nanoparticles. Also highlighted are the cellular pathways modulated by cells for survival and the exemplified cross-toxicities that can occur between copper-based nanoparticles and pharmacological agents.
.

## INTRODUCTION

Manufactured nanoparticles are more and more widely used in more and more consumer products, ranging from personal care products to tyres and concrete. Among the nanoparticles, metals and metal oxides represent an important part of the total production and are used in water treatment, as antibacterials, in antifouling paints or in microelectronics. These wide uses pose in turn the problem of the toxicological evaluation of the toxicity of these nanoparticles (1, 2), and especially of the long-term effects that often do not come from simple cell mortality but from altered cellular functions.

Macrophages are one of the cell types that deserve special attention in toxicology, due to the variety of their functions. An altered cytokine production can lead to adverse long term effects, as documented for example in the case of asbestos (3). Other dysfunctions of the innate immune system can lead to deregulation of the immune responses and to severe adverse effects, e.g. a higher incidence of tumours (4).

It is therefore not surprising that immunotoxicology of nanoparticles is a developing field (5-7), and several studies have been devoted to the response of macrophages to nanoparticles. However, most of these studies are limited to the effect of nanoparticles on cell viability and on cytokine production (e.g. (8-11)) although some publications also study oxidative stress (12-14) and sometimes other functional parameters (15-17) . Very few studies use the analytical power of proteomics to go deeper into the mechanisms of response to nanoparticles or to metals (reviewed in (18)). A few exceptions exist, e.g. on carbon-based nanoparticles (19) and on titanium dioxide (20, 21).

Most of the toxicological studies in this field have been focused on a few nanoparticles, either used as health products such as iron oxide (15, 17, 22), or used in variety of consumer products such as silver (13, 14), silica (9, 12) or titanium dioxide (11, 16, 20, 21).

However, many other nanoparticles are more and more used in industrial applications without extensive toxicological testing. Good examples are indium-tin oxide, used in electronic screens and which appears to be toxic (23) and the copper-based nanoparticles, used in high-performance batteries (24), in water depollution (25), and as bactericidal in replacement of nano-silver. Copper and copper oxide induce a strong toxicity (26, 27), coupled with inflammation (28), oxidative stress (29) and genotoxicity (30), at least in epithelial cells.

In the light of these various effects, we have decided to use a combination of a proteomics approach with targeted approaches to address in more molecular details the responses of macrophages to copper-based nanoparticles, i.e. both metallic copper and copper II oxide.

## MATERIAL AND METHODS

### Nanoparticles

Metallic copper and copper oxide nanoparticles (<50 nm) were purchased from Sigma-Aldrich (catalog numbers 684007 and 544868 respectively). They were dispersed in water as a 5.5% w/v suspension by sonication for 60 minutes in a cup-horn instrument (BioBlock Scientific, France) , under a 5°C thermostated water circulation. One-tenth volume of 10% w/v PVP40 solution was added under sterile conditions, and the particles were coated for 1 hour under constant agitation. The actual size of the particles was determined after dilution in water or in complete culture medium by dynamic light scattering, using a Wyatt Dynapro Nanostar instrument or a Malvern HS 3000 instrument, the latter instrument being also used to determine the zeta potential.
The morphology of the samples was observed by SEM (Scanning Electron Microscopy). A 200 mesh carbon grid was dipped in the nanoparticles suspensions and dried under air before imaging.
The amount of coating attached on the inorganic nanoparticles was evaluated from weight loss (from about 10mg of sample) after annealing under air using a Thermogravimetric Analysis device (Setaram). The temperature cycle was heating at a rate of 10°C/min up to 600°C followed by a dwelling time of 30 minutes and natural cooling.
Zirconium oxide nanoparticles (<100 nm) were purchased from Sigma-Aldrich as a 10% (w/v) dispersion in water (catalog number 643025). Prior to use, they were diluted by mixing one volume of dispersion with one volume of 2% w/v PVP-40 for 1 hour under constant agitation. The actual size of the final dispersion in complete culture medium was determined as for the copper nanoparticles.

## Nanoparticles dissolution in culture medium

Nanoparticles were added at 5, 10 or 20 µg/ml to complete culture medium (RPMI 1640 + 10% fetal bovine serum) in cell culture 6 well plates containing 2ml culture medium per well. In some experiments, conditioned medium, i.e. complete cell culture medium having been in contact with the cells for at least 24 hours, was used in place of fresh complete culture medium. Known concentrations of copper chloride were also added to complete culture medium in control wells and incubated under the same conditions. The plates were incubated for 24 hours in a cell culture incubator at 37°C and 5% $CO_2$ atmosphere. The culture medium was recovered, and centrifuged at 270,000g for 45 minutes to sediment the nanoparticles (31). The concentration of copper ions was then determined using the Zincon method (32). Briefly, 1 ml of supernatant was acidified with trichloroacetic acid (7.5% w/v final concentration) to precipitate proteins and release bound copper ions. This precipitation step was carried out for 30 minutes on ice. The precipitated proteins were eliminated by centrifugation (15,000 g, 15 minutes), and the supernatants were collected. After neutralization by addition of 0.3 volumes of 2 M Tris, Zincon was added (50 µM final concentration) and the copper concentration was determined by absorbance at 600 nm, using the copper chloride in medium points to build the standard curve.

## Cell culture

### Cell lines

The mouse macrophage cell line RAW 264.7 was obtained from the European Cell Culture Collection (Salisbury, UK). The cells were cultured in RPMI 1640 medium +

10% fetal bovine serum. Cells were seeded every two days at 200,000 cells/ml and harvested at 1,000,000 cells per ml.

Primary macrophages

Primary mouse macrophages were obtained from bone marrows essentially as described by Schleicher and Bogdan (33) and Tarakanova et al. (34) . Briefly, bone marrow was aseptically collected from femur and tibias of 4 weeks-old C57Bl6 mice, sacrificed by cervical dislocation. The marrow plugs were dissociated by repeated flushing in the culture medium (DMEM 65% v/v, L929 cell culture supernatant 20% v/v, fetal bovine serum 10% v/v and horse serum 5% v/v) and plated at 4 million cells/ml on bacteriology culture dishes. An equal volume of medium was added after 4 days in culture. After 7 days of culture, the cells were scraped and replated at 1 million cells/ml in the same medium, except that the L929 supernatant was lowered to 10% v/v and DMEM increased to 75% v/v. Mature macrophages were obtained after an additional 3 days, and were viable and stable up to 2 weeks, provided that the culture medium was renewed every 3 days.

Cell treatment and viability assessment

Cells were treated with variable concentrations of nanoparticles, added directly in the culture medium. After 24 hours in the presence of the nanoparticles or ions, cell viability was measured by trypan blue exclusion. For treatment with inhibitors, cells were pretreated for 6 hours with either 100µM buthionine sulfoximine or 200 nM rotenone or 200 nM antimycin A. Copper oxide nanoparticles (or copper chloride) were then added for a further 18 hours before collection of the cells and viability measurements.

Neutral red uptake assay

This assay was performed essentially as described in (35). Briefly, cells grown in 12-well plates and treated or not with nanoparticles were incubated for 1 hour with 40 µg/ml neutral red (final concentration, added from a 100x stock solution in 50% ethanol water). At the end of the incubation period, the medium was discarded and the cell layer was rinsed twice with PBS (5 minutes per rinse). The PBS was removed, and 1 ml of elution solution (50% ethanol and 1% acetic acid in water) was added per well. The dye was eluted for 15 minutes under constant agitation, and the dye concentration was read spectrophotometrically at 540 nm. The results were expressed as % of the control cells (untreated with nanoparticles).

Phagocytosis activity measurement

The phagocytic acitivity was measured using fluorescent latex beads (1µm diameter, green labelled, catalog number L4655 from Sigma), essentially as described in (36). Briefly, the beads were pre-incubated at a final concentration of 55 mg/ml for 30 minutes at 37°C in a 1:1 (v/v) PBS/ horse serum solution. They were then incubated with the cells (5.5 mg/L final concentration) for 2 h at 37°C. The cells were then harvested, washed twice with cold PBS, and incubated for 5 minutes on ice with a 0.25 M D-sorbitol solution in order to remove beads adhering to the cell surface. After another wash with cold PBS, cells were analyzed by flow cytometry on a FACScalibur instrument (Beckton Dickinson), using fluorescein excitation/emission filters, and using the live cell selection as described above.

Transmission Electron Microscopy

For TEM, fixation was initiated by adding an equal volume of fixative solution, previously warmed to 37°C, to the cells after treatment with copper oxide nanoparticles for 24 hours. The fixative solution contained 5% glutaraldehyde (Electron Microscopy Sciences, Euromedex, Strasbourg, France) in a 0.1 M sodium cacodylate buffer (both Merck, Darmstadt, Germany) (305 mOsm, pH 7.3). After 10 min the mixture was centrifuged, the supernatant was discarded, and the pellet resuspended in the fixative solution containing 2.5% glutaraldehyde in 0.1 M sodium cacodylate buffer for 45 min at room temperature. The cells were then washed in 0.1 M sodium cacodylate buffer and postfixed for 1 h at 4°C with 1% osmium tetroxide (Merck) in the same buffer and stained for 1 h at 4°C in 4% uranyl acetate. After further washing in distiled water, the cells were dehydrated in graded (50, 70, 80, 95, and 100%) ethanol solutions, incubated for 1 h in Epon (Electron Microscopy Sciences): absolute alcohol (1:1, vol/vol), then overnight in Epon and embedded in Epon. Ultrathin sections, stained with lead citrate (*Leica*, Bron, France) and uranyl acetate (Merck), were examined under a Philips CM 120 BioTwin electron microscope (120 kV).

Intracellular glutathione measurements

Intracellular glutathione levels were analyzed by the monochlorobimane technique (37), with some modifications (38). Briefly, the cells were harvested, centrifuged for 5 minutes and then labelled for 5 minutes with a 50 µM monochlorobimane solution diluted in PBS at 37°C. The reaction was stopped for 5 minutes on ice in the dark, and the cells were then washed twice with cold PBS. Finally, the cells were analyzed by flow cytometry on a MoFlo instrument (Beckman Coulter) using a UV laser for excitation and reading the emission at 450 nm. The analysis was restricted to live cells only.

Mitochondrial transmembrane potential assessment

The mitochondrial transmembrane potential was assessed by Rhodamine 123 uptake (39). Briefly, cells were treated with 10 µg/ml Rhodamine 123 in complete culture medium for 30 minutes at 37 °C. Cells were then harvested, rinsed twice in cold PBS and analysed by flow cytometry on a FACScalibur instrument (Beckton Dickinson), using fluorescein excitation/emission filters. First the live cells were selected on the basis of the size and granularity, then the fluorescence of these live cells was measured using the fluorescein channel.

NO production

The cells were grown to confluence in a 6 well plate, and then treated or not with copper nanoparticles, copper oxide nanoparticles, or copper chloride. After 24 hours of treatment, half of the wells were treated with 1 µg/ml LPS (from salmonella), and arginine monohydrochloride was added to all the wells (5 mM final concentration) to give a high concentration of substrate for the nitric oxide synthases. After 24 hours of incubation, the cell culture medium was recovered, centrifuged at 10,000 g for 10 minutes to remove cells and debris, and the nitrite concentration was read at 540 nm after addition of an equal volume of Griess reagent and incubation at room temperature for 30 minutes

## Proteomics

### 2D gel electrophoresis

#### Sample preparation

The cells were collected by scraping, and then washed three times in PBS. The cells were then washed once in TSE buffer (Tris-HCl 10 mM pH 7.5, sucrose 0.25M, EDTA 1 mM), and the volume of the cell pellet was estimated. The pellet was resuspended in its own volume of TSE buffer. Then 4 volumes (respective to the cell suspension just prepared) of concentrated lysis buffer (urea 8.75 M, thiourea 2.5 M, CHAPS 5% w/v, TCEP-HCl, 6.25 mM, spermine base 12.5 mM) were added and the solution was let to extract at room temperature for 1 hour. The nucleic acids were then pelleted by ultracentrifugation (270,000 g at room temperature for 1 h), and the protein concentration in the supernatant was determined by a dye-binding assay (40). Carrier ampholytes (Pharmalytes pH 3-10) were added to a final concentration of 0.4% (w/v), and the samples were kept frozen at -20°C until use

#### Isoelectric focusing

Home made 160 mm long 4-8 linear pH gradient gels (41) were cast according to published procedures (42). Four mm-wide strips were cut, and rehydrated overnight with the sample, diluted in a final volume of 0.6 ml of rehydration solution (7 M urea, 2 M thiourea, 4% CHAPS, 0.4% carrier ampholytes (Pharmalytes 3-10) and 100mM dithiodiethanol (43)(44).
The strips were then placed in a Multiphor plate (GE Healthcare), and IEF was carried out with the following electrical parameters: 100V for 1 hour, then 300V for 3 hours, then 1000V for 1 hour, then 3400 V up to 60-70 kVh. After IEF, the gels were equilibrated for 20 minutes in Tris 125mM, HCl 100mM, SDS 2.5%, glycerol 30% and urea 6 M (45) . They were then transferred on top of the SDS gels and sealed in place with 1% agarose dissolved in Tris 125mM, HCl 100mM, SDS 0.4% and 0.005% (w/v) bromophenol blue.

#### SDS electrophoresis and protein detection

Ten percent gels (160x200x1.5 mm) were used for protein separation. The Tris taurine buffer system (46) was used and operated at a ionic strength of 0.1 and a pH of 7.9. The final gel composition is thus Tris 180mM, HCl 100 mM, acrylamide 10% (w/v), bisacrylamide 0.27%. The upper electrode buffer is Tris 50 mM, Taurine 200 mM, SDS 0.1%. The lower electrode buffer is Tris 50 mM, glycine 200 mM, SDS 0.1%. The gels were run at 25V for 1hour, then 12.5W per gel until the dye front has reached the bottom of the gel. Detection was carried out by fast silver staining (47).

#### Image analysis

The gels were scanned after silver staining on a flatbed scanner ( Epson perfection V750), using a 16 bits grayscale image acquisition. The gel images were then analyzed using the Delta 2D software (v 3.6). 4 gels coming from 4 independent cultures were used for each experimental group. Spots that were never expressed above 100 ppm of the total spots were first filtered out. Then, significantly-vazrying spots were selected on the basis of their Student T-test p-value between the treated and the control groups. Spots showing a p-value lower than 0.05 were selected. However, two additional steps of statistical validation were used. On a global point perspective, the Storey and Tibshirani approach (48) was used to evaluate a FDR, as described previously (49). Then each of the selected spots was tested for significance by the Mann-Whitney ranking test, to rule out any bias that could exist the the T-test due to its underlying hypotheses.

Mass spectrometry

The spots selected for identification were excised from silver-stained gels and destained with ferricyanide/thiosulfate on the same day as silver staining in order to improve the efficiency of the identification process (50) (51). In gel digestion was performed with an automated protein digestion system, MassPrep Station (Waters, Milford, USA). The gel plugs were washed twice with 50 µL of 25 mM ammonium hydrogen carbonate ($NH_4HCO_3$) and 50 µL of acetonitrile. The cysteine residues were reduced by 50 µL of 10 mM dithiothreitol at 57°C and alkylated by 50 µL of 55 mM iodoacetamide. After dehydration with acetonitrile, the proteins were cleaved in gel with 10 µL of 12.5 ng/µL of modified porcine trypsin (Promega, Madison, WI, USA) in 25 mM $NH_4HCO_3$. The digestion was performed overnight at room temperature. The generated peptides were extracted with 30 µL of 60% acetonitrile in 0.1% formic acid. Acetonitrile was evaporated under vacuum before nanoLC-MS/MS analysis.
NanoLC-MS/MS analysis was performed using on nanoLC-QTOF-MS system and on nanoLC-IT-MS system. The nanoLC-QTOF-MS system was composed of the nanoACQUITY Ultra-Performance-LC (Waters Corporation, Milford, USA) coupled to the Synapt™ High Definition Mass Spectrometer™ (Waters Corporation, Milford, USA). The system was fully controlled by MassLynx 4.1 SCN639 (Waters Corporation, Milford, USA).
The nanoLC system was composed of ACQUITY UPLC® BEH130 C18 column (250 mm x 75 µm with a 1.7 µm particle size, Waters Corporation, Milford, USA) and a Symmetry C18 precolumn (20 mm × 180 µm with a 5 µm particle size, Waters Corporation, Milford, USA). The solvent system consisted of 0.1% formic acid in water (solvent A) and 0.1% formic acid in acetonitrile (solvent B). 4 µL of sample were loaded into the enrichment column during 3 min at 5 µL/min with 99% of solvent A and 1% of solvent B. Elution of the peptides was performed at a flow rate of 300 nL/min with a 8-35% linear gradient of solvent B in 9 minutes.
The tandem mass spectrometer was equipped with a Z-spray ion source and a lock mass system. The capillary voltage was set at 2.8 kV and the cone voltage at 35 V. Mass calibration of the TOF was achieved using fragment ions from Glu-fibrino-peptide B on the [50;2000] m/z range. Online correction of this calibration was performed with Glu-fibrino-peptide B as the lock-mass. The ion $(M+2H)^{2+}$ at m/z 785.8426 was used to calibrate MS data and the fragment ion $(M+H)^{+}$ at m/z 684.3469 was used to calibrate MS/MS data during the analysis. The system was operated in Data-Dependent-Acquisition (DDA) mode with automatic switching between MS (0.5 s/scan on m/z range [150;1700]) and MS/MS modes (0.5 s/scan on m/z range [50;2000]). The two most abundant peptides (intensity threshold 20 counts/s), preferably doubly and triply charged ions, were selected on each MS spectrum for further isolation and CID fragmentation using collision energy profile. Fragmentation was performed using argon as the collision gas.
Mass data collected during analysis were processed and converted into .pkl files using ProteinLynx Global Server 2.3 (Waters Corporation, Milford, USA). Normal background subtraction type was used for both MS and MS/MS with 5% threshold and polynomial correction of order 5. Smoothing was performed on MS/MS spectra (Savitsky-Golay, 2 iterations, window of 3 channels). Deisotoping was applied for both MS (medium desisotoping) and MS/MS (fast deisotoping).

The nanoLC-IT-MS system was composed of the Agilent 1200 series nanoLC-Chip system (Agilent Technologies, Palo Alto, USA) coupled to the amaZon ion trap (Bruker Daltonics GmbH, Bremen, Germany). The system was fully controlled by HyStar 3.2 and trapControl 7.0 (Bruker Daltonics, Bremen, Germany).
The chip was composed of a Zorbax 300SB-C18 (43 mm × 75 µm, with a 5µm particle size) analytical column and a Zorbax 300SB-C18 (40 nL, 5 µm) enrichment column. The solvent system consisted of 2% acetonitrile, 0.1% formic acid in water (solvent A) and 2% water, 0.1% formic acid in acetonitrile (solvent B). 4 µL of sample were loaded into the enrichment column at a flow rate set to 3.75L/min with solvent A. Elution of the peptides was performed at a flow rate of 300 nL/min with a 8-40% linear gradient of solvent B in 7 minutes.
For tandem MS experiments, the system was operated in Data-Dependent-Acquisition (DDA) mode with automatic switching between MS and MS/MS. The voltage applied to the capillary cap was optimized to -1850V. The MS scanning was performed in the standard enhanced resolution mode at a scan rate of 8100 m/z per second. The mass range was 250-1500 m/z. The Ion Charge Control was 200000 and the maximum accumulation time was 200 ms. A total of 2 scans was averaged to obtain a MS spectrum and the rolling average was 1. The six most abundant precursor ions with an isolation width of 4 m/z were selected on each MS spectrum for further isolation and fragmentation. The MS/MS scanning was performed in the ultrascan mode at a scan rate of 32500 m/z per second. The mass range was 100-2000 m/z. The Ion Charge Control was 300000. A total of 2 scans was averaged to obtain an MS/MS spectrum.
Mass data collected during analysis were processed and converted into .mgf files using DataAnalysis 4.0 (Bruker Daltonics GmbH, Bremen, Germany). A maximum number of 1700 compounds was detected with an intensity threshold of 150000. A charge deconvolution was applied on the MS full scan and the MS/MS spectra with an abundance cutoff of 5% and 2% respectively and with a maximum charge state of 3 and 2 respectively.
For protein identification, the MS/MS data were interpreted using a local Mascot server with MASCOT 2.4.0 algorithm (Matrix Science, London, UK) against UniProtKB/SwissProt (version 2012_08, 537,505 sequences). The research was carried out in all species. A maximum of one trypsin missed cleavage was allowed. Spectra from Qtof were searched with a mass tolerance of 15 ppm for MS and 0.05 Da for MS/MS data and spectra from Ion Trap were searched with a mass tolerance of 0.3 Da in MS and MS/MS modes. Carbamidomethylation of cysteine residues and oxidation of methionine residues were specified as variable modifications. Protein identifications were validated with at least two peptides with Mascot ion score above 20 or with one peptide with Mascot ion score above 30 and 5 consecutive fragment ions in MS/MS spectrum.

## RESULTS AND DISCUSSION

### Nanoparticle characterisation in culture medium

Upon sonication, copper and copper oxide nanoparticles could be dispersed to a microaggregate size of ca. 250 nm. These microaggregates were stable upon dilution in pure water, but we observed an immediate aggregation over the micron size when the nanoparticles were diluted in culture medium, even when containing fetal bovine serum. Unfortunately, this aggregation resulted in immediate sedimentation, leading to a very uneven exposure of the cell layer to the particles in the cell culture dish.

All of the undesirable phenomena were prevented when the copper and copper oxide nanoparticles were coated with polyvinylpyrrolidone (PVP) prior to dilution in the cell culture medium. PVP has been used in the preparation of stable copper nanoparticles by aqueous routes (52-54), e.g., for conductive inks (55). PVP is also used for the stabilisation of silver nanoparticles (56, 57) and in commercial nanoparticle dispersions (e.g., silver nanoparticles from Buhler). It is also classified as a biocompatible polymer (58, 59) and does not activate the macrophages by itself (60). Indeed, it induced no biological response, neither in our proteomic experiments nor in the targeted ones (data not shown).

Light scattering and scanning electron microscopy experiments showed that coating with PVP did not significantly change the microaggregation in water. However, the macroaggregation observed in culture media for noncoated particles was prevented, and the average diameter was kept close to 250 nm (Figure 1). The coating process with PVP significantly changed the zeta potential, measured at 0 mV instead of the classical 30 mV of a water suspension of CuO (61).

Zirconium oxide nanoparticles had the same aggregation property in culture media, which was prevented as well by PVP coating. The resulting average size was 180 nm measured by light scattering.

Dissolution studies, carried out in complete cell culture medium and in conditioned cell culture medium (both without cells), revealed that ca. 60% of the copper present initially in the nanoparticles was dissolved in the culture medium, and that dissolution was linear in the 5–20 µg/ml range. The dissolution data were similar whether fresh or conditioned medium was used and whether uncoated or PVP-coated nanoparticles were used. These dissolution figures are quite close to those reported recently under the same conditions, i.e., dispersion in water then dilution in complete cell culture medium (31).

### Determination of the effective copper doses

For carrying out a proteomics study, it was necessary to determine the dose to obtain the best compromise between viability and biological effect. We, therefore, decided to use an LD20, which is a dose leading to a cell mortality of 20%, knowing that the mortality of the control culture is approximately 5%. As nanoparticles are known to interfere with many viability estimation methods (62), we chose a very conservative and robust dye exclusion method for this critical step (63).

The results show that the LD20 was obtained at 10 µg/ml for the Cu and CuO nanoparticles for the RAW264 cell line, while it was obtained at 5 µg/ml for primary macrophages (Figure 2). A copper ion concentration of 0.2 mM corresponded to the LD20 for primary macrophages, while it induced no more than 10% mortality for the cell lines.

We, therefore, fixed our working concentrations at LD20 for the nanoparticles and at 0.2 mM for the copper ions.

## Determination of nanoparticle fate in cells

In order to determine the fate of the nanoparticles in the presence of macrophages, transmission electron microscopy was performed on ultrathin sections of cells. These experiments showed that nanoparticles were present in vacuolar structures that were identified as multivesicular bodies and as lysosomes (Figure 3). These results were completely in line with what can be expected from professional phagocytic cells, as macrophages are. Moreover, these results demonstrate that macrophages seem to ingest nanoparticle aggregates present in the culture medium and to break them down within the endosomes-lysosomes. The first step was obviously the disaggregation of the aggregates into the nominal nanoparticles. However, the disaggregation did not seem to stop at this stage, as much smaller copper oxide particles were present (Figure 3). This observation is consistent with an intracellular dissolution of the copper oxide nanoparticles, which could be expected in an acidic organelle such as the lysosome. Macrophages are known to be able to uncoat nanoparticles (17), and this is further confirmed in our case by the existence of eroded nanoparticles within the cells. This size distribution was observed at both 4 hours and 24 hours of treatment with the nanoparticles. Thus, it can be concluded that the intracellular solubilisation of nanoparticles, which seems important for their biological effects (64), is kept even with coated nanoparticles. Overall, the coating process does not seem to interfere either with the extracellular or with the intracellular processes.

## Proteomic analyses

The proteomic analyses were carried out on the RAW264 cell line, and typical gels are shown in Figure 4. More than 1,400 reliable spots were detected by the image analysis software, and they were further analysed for quantitative variations. As for all omics strategies, the issue of multiple testing must be taken into account. In other terms, positive hits will always be obtained when many molecules are screened simultaneously, as in many comparative omic screens, just because of the statistical noise. Most classical statistical corrections for multiple testing, such as the Bonferroni correction, are not suited to omics data, where relatively low replicate numbers are available, and are in fact much too conservative (65). We, therefore, used the t-test distribution approach, pioneered by Storey and Tibshirani (48) for transcriptomics and first applied to proteomics by Karp and coworkers (49). The distribution of the p-values (*t*-test) for the three tested conditions (nanoparticular Cu, nanoparticular CuO and Cu ion) is shown in Supplementary figure 1. This approach showed us that we could not expect any really significant result with a 25-µM dose of copper ion, which has no effect on the cells. However, significant results could be expected for a higher dose of copper ion, corresponding to the LD20, as well as for copper and even more for copper oxide nanoparticles at the LD20.

We, therefore, focused our analysis on spots that showed a consistent variation between the Cu and CuO treatment, and these spots are indicated in Figure 4. These spots were then identified by mass spectrometry, and the results are shown in Table 1, with more details in Supplementary Tables 1 and 2. Some neighbouring spots or spots corresponding to other proteins of interest were also identified. On a global level, in addition to the classical metabolic modulation (66, 67), some cellular compartments/functions were highlighted by this proteomic analysis, such as cytoskeleton, cell signalling, mitochondrion and oxidative stress response, especially in the glutathione system.

In order to check for the specificity of the proteomic variations observed, we used zirconium oxide nanoparticles. These particles show very weak cytotoxicity (68) and were indeed nontoxic to RAW 264 cells up to 100 µg/ml. As the density of zirconium oxide (5.68) is close to that of copper oxide (6.31), it meant that treatment of the cells with an equal concentration in weight also corresponded to equal concentrations in particle numbers, which is an important parameter for phagocytic cells. Furthermore, as zirconium oxide nanoparticles are also dispersed with PVP, this control integrated any possible effect of absorbed PVP.

Treatment with 10 µg/ml zirconium oxide for 24 hours induced very few changes detectable in our proteomic screen (Supplementary Table 3 and Supplementary Figure 2), and none of these changes corresponded to the ones induced by copper and copper oxide nanoparticles (Table 1). This showed the specificity of our proteomic results. We, therefore, decided to test specifically these functions by targeted analyses, both on the RAW 264 cell line and on primary macrophages derived from bone marrow. As the proteomic effects were more pronounced with CuO nanoparticles, we primarily used these nanoparticles for the targeted experiments; although some experiments were also carried out with copper nanoparticles.

Characterisation of the general stress response

Quite frequently, proteomic studies identify a general cellular stress response module, which encompasses proteins involved in central metabolism and protein production and folding (67). However, in the case of the response of macrophages to copper-based nanoparticles, we did not observe the classical induction in central metabolism proteins. In contrast, we observed a small decrease in malate dehydrogenase, galactokinase and lactate dehydrogenase. We also did not observe changes in the heat shock proteins, in contrast to what has been described at the mRNA level in lung cells exposed to copper oxide nanoparticles (31), or at the protein level for cells exposed to gold nanoparticles (69). However, and consistent with the latter publication, we also observed a strong induction of the ribosomal protein P0.

Characterisation of the oxidative stress response

Oxidative stress has been observed in many cell types and with many nanoparticles [reviewed for example in (70)]. However, cellular responses to oxidative stress are not well characterised. The cellular oxidative stress response system is quite complex and involves several classes of enzymes, including catalases, peroxidases (including peroxiredoxins), superoxide dismutases and enzymes producing and using glutathione.

Our proteomic screen revealed that the heme oxygenase was very strongly induced, with a complex pattern of modifications. For the more classically observed peroxiredoxins (66, 67), we could determine by using the BHP oxidation test (71) that the induced form of Prx1 is indeed the oxidized form, while the native form of Prx6 is induced in our experiments. Other peroxiredoxins, such as Prx2 and the mitochondrial Prx3, remained unchanged, as shown in Table 1. The induction of Prx1 in response to gold nanoparticles has been described recently (69), but the status of the other peroxiredoxins had not been investigated. In addition, Prx1 is known to be inducible by heavy metals, as is heme oxygenase (72).

Furthermore, our results are consistent with those obtained for Prx6 with oxidative stress in lung cells (73) and in myoblasts (74).

As to the superoxide dismutases, we observed a decrease in the copper-zinc superoxide dismutase. The response of copper-zinc superoxide dismutase seems to be variable, because either an increase (75), decrease (76) or no change (29) has been described in response to metallic nanoparticles. We also observed no change for the mitochondrial manganese superoxide dismutase, although an increase in manganese superoxide dismutase has been described at the mRNA level (not at the protein level) in lung cells exposed to copper oxide nanoparticles (31).

Finally, as to the glutathione-dependent oxidative stress response system, we observed an increase in formyl glutathione hydrolase and a strong induction of the regulatory subunit of the glutamate-cysteine ligase, i.e., the activating subunit of the limiting enzyme in glutathione biosynthesis (77). This suggested that the cells responding to copper-based nanoparticles make a strong effort to increase their glutathione level, and we tested the intracellular levels of reduced glutathione by a chlorobimane conjugation test (37). Compared with the positive control represented with diethyl maleate, which destroys reduced glutathione by alkylation, copper ion and copper-based nanoparticles induced a more subtle effect (Figure 5). A subpopulation of cells with low levels of reduced glutathione was induced, especially by copper nanoparticles. However, most of the cells had normal reduced glutathione levels, so that the average for the whole cell population was not significantly changed in most cases. In addition, the glutathione reductase was not induced by copper treatment (Figure 5).

Primary macrophages, however, were more sensitive and showed a decrease in their free glutahione levels of copper ions and of the two-copper-based nanoparticles. These results were consistent with those observed previously with copper oxide nanoparticles in lung cells (29, 30) and with silver in liver cells (75, 78). In the case of copper-based nanoparticles, this is also consistent with the well-known role of glutathione as a copper buffer in cells (79). Consistent with this role, the decrease in free glutathione levels was also observed when the cells were treated with copper ions.

However, the RAW264.7 cells appeared less sensitive to copper oxide and copper ions, as if the strong induction of glutamate-cysteine ligase was sufficient in this case to restore close to normal reduced glutathione levels. To verify this hypothesis, we treated RAW264.7 cells with non toxic doses of buthionine sulfoximine, an inhibitor of glutamate-cysteine ligase. This induced a marked hypersensitivity of the cells to copper-based nanoparticles, as shown on Figure 5 (panel F)

Overall, these proteomics-based results show the fine tuning of the cellular response to oxidative stress, as exemplified from the differential response of the three cytosolic peroxidoxins (Prx1, 2 and 6), and from the response of the glutathione-dependent system.

Characterisation of the mitochondrial response

Mitochondrial function is sometimes tested in response to nanoparticle exposure by testing the main parameter of mitochondrial transmembrane potential. A decrease in mitochondrial transmembrane potential has been observed in response to copper-based nanoparticles or to copper ions in liver (78), in lung cells (80) and in macrophages (81). We also observed such a decrease in our study using a Rhodamine 123 uptake test (39), as shown in Figure 6. Both nanoparticles and copper ions induced a dose-dependent decrease in the mitochondrial transmembrane potential, and this effect was more pronounced in the primary macrophages than in the cell line.

Little is known, however, about the molecular events that could be related to the observed decrease in the mitochondrial transmembrane potential, which could even be due to a change in the plasma membrane potential. However, our proteomics screen suggested some possible mechanisms. First of all, we did not observe any change in the two major mitochondrial antioxidant proteins, superoxide dismutase 2 and peroxiredoxin 3. This suggests that there is no massive mitochondrial oxidative stress. However, we observed an increase in the levels of mit-EF-Tu. As this protein is a translation factor for the intramitochondrial synthesis of the few respiratory chain proteins encoded by the mitochondrial genome, this result suggested an increase in the synthesis of the proteins of the oxidative phosphorylation chain, corroborated by an increase in one nuclear-coded subunit of the NADH dehydrogenase and one of the ubiquinol-cytochrome c oxidoreductase complex. However, the synthesis of another subunit of NADH dehydrogenase containing an iron-sulfur cluster (Ndufs8) was decreased. This suggests a toxic mechanism in which copper ions pumped into the mitochondria may alter the function of redox-sensitive proteins containing iron-sulfur clusters and, thus, alter both electron transfer and proton pumping through the inner mitochondrial membrane, leading to the observed decrease in the transmembrane potential. In this context, the observed increases in mit-EF-Tu, Uqcrc1 and Ndufs3 would represent an attempt of the cells to restore a normal mitochondrial transmembrane potential, which is observed at nontoxic doses of nanoparticles in RAW 264 cells. In order to test this hypothesis, we analysed cell survival after treatment with subtoxic doses of inhibitors targeting the two respiratory complexes highlighted by proteomics, namely Complex I and Complex III, inhibited respectively by rotenone and antimycin A. The results show the synergistic toxicity of mitochondrial inhibitors and copper oxide nanoparticles (Figure 6).

Changes in the cytoskeleton, intracellular trafficking and phagocytosis

Our proteomic screen pointed to changes in the cytoskeleton induced by copper-based nanoparticles. However, not all types of cytoskeletal proteins were altered by the nanoparticles. The tubulin cytoskeleton did not show any change in our screen, and besides vimentin, mostly the actin-myosin cytoskeleton was altered, either directly (decrease in tropomyosins) or indirectly (decrease in Rho GDI inhibitors, which is functionally equivalent to an increase in the activity of the Rho proteins). Rho proteins control specific modulations of the actin cytoskeleton (e.g., formation of stress fibres) that have been associated with decreased phagocytic capacity in macrophages stressed by hyperoxia (82).

A decrease in a subunit of the proton pump also pointed to a possible phagocytic deficiency. This prompted us to assay phagocytosis, which is a key macrophage function. CuO nanoparticles had no effect at moderately toxic doses in cell lines, and an inhibition of phagocytosis was apparent only at a toxic dose (20 µg/ml) (Figure 7). Cu nanoparticles were more efficient at inhibiting phagocytosis, while copper ions had no effect. In primary macrophages, which are more sensitive to nanoparticles in all our tests, both Cu and CuO nanoparticles were efficient in inhibiting phagocytosis, while Cu ions had no effect. Interestingly, tropomyosin, Rho GDI inhibitors and proton pump subunits were not modulated by copper ions in our proteomic screen.

To further study the functional effects of the observed decrease in the proton pump, we performed a classical neutral red uptake assay. Although classically used as a general viability assay (35), neutral red uptake is indeed an assay of lysosomal function because the acidic conditions retain neutral red inside the lysosomes.

Figure 7C shows that neutral red uptake decreases more severely upon treatment with copper oxide nanoparticles than should be expected from a simple decrease in cell viability, showing a specific effect on lysosomes and validating the proteomically observed decrease in the proton pump.

## NO production

In addition to phagocytosis, production of cytokines and of NO is a key function of macrophages. We, therefore, tested the production of NO in two different setups. In the first setup, we tested if nanoparticles could directly induce the production of NO, as this has been shown to occur with silver nanoparticles (13). In the second setup, we tested the modulation of the LPS-induced production of NO by a previous treatment with nanoparticles (Figure 8). While Cu-NP, CuO-NP and copper ions did not induce the production of NO by themselves, all three moderately inhibited LPS-induced NO production in primary macrophages. However, only copper ions were able to reduce the LPS-induced NO production in the case of the RAW 264 cell line. These results on NO production could be linked, at least indirectly, with the decrease in S-Adenosylhomocysteine hydrolase observed in our proteomic screen. Inhibition of S-Adenosylhomocysteine hydrolase has been shown to reduce pro-inflammatory properties (83), and this can be linked, at least in part to a decrease in adenosine concentrations, as adenosine is a potent activator of nitric oxide production in macrophages (84). Through this pathway, a reduced level of S-Adenosylhomocysteine hydrolase would suggest a lower ability to produce NO, at least in response to LPS, and this is exactly what was observed when NO production was specifically tested.

## Overall, what can be explained by copper ions in the toxicity of nanoparticles ?

Since the demonstration of an important copper ion release from copper-based nanoparticles in biological media (85), it is generally thought that most of the nanoparticle toxicity arises from the copper ion itself, and not directly from the nanoparticles. However, transcriptomic studies (31), as well as our work presented here, show that extracellular copper ions can only account for a part of the cellular responses to copper-based nanoparticles. In our case, extracellular copper ions induced the general cellular stress responses (mitochondrial or glutathione dependent) and altered some specific cellular responses (e.g., nitric oxide production) but not all. For example, phagocytosis, a key function of macrophages, is altered by the nanoparticles but not by the free ions. We could, however, rule out a mechanical explanation of this inhibition, as titanium dioxide nanoparticles do not inhibit phagocytosis (data not shown).

These differences between the responses to the ion itself and to the nanoparticles may be the result of the existence of two different copper channels at the plasma membrane (Ctr1) and at the lysosomal membrane (Ctr2) (86). Thus, the real intracellular concentration, and even more subtly, the local copper concentrations within the cell can differ depending on the mode of entry. This might be the case for highly phagocytic cells such as macrophages, for which the lysosome-dependent entry can be much more important than the plasma membrane-dependent entry.

## Concluding remarks

Despite its limited depth, estimated to ca. 500 proteins [1,500 spots for an average of 3 spots per protein (87)] the proteomics screen used here allowed us to point out several cellular functions that were modified upon treatment of macrophages with copper-based nanoparticles and/or copper ions. Interestingly, copper treatment was associated with a reduction in protein levels, suggesting either a decrease in their synthesis or an increase in their degradation. The fact that some copper-binding proteins (e.g., copper-zinc superoxide dismutase and S-adenosyl homocysteine hydrolase) diminished suggests that an excess in intracellular copper could lead to protein destructuration and then degradation.

Moreover, the increase in the glutamate cysteine ligase subunit and of some mitochondrial proteins shows how proteomics can point to cellular resistance mechanisms even when they are successful, so that no apparent change can be detected in the downstream parameter (e.g., glutathione levels).

In the frame of macrophage physiology, the uptake of copper does not bring an activation of the macrophages, with the associated increase in the production of inflammatory mediators, as described for SiO2 an TiO2 nanoparticles (4) (88). It rather brings a depression of core macrophage activities such as NO production and phagocytosis, which can have in turn consequences on the overall efficiency of the immune system.

Finally, our results also show that even at nonlethal doses, nanoparticles can act synergistically with other toxicants to induce cell mortality and dysfunction.

## AUTHOR CONTRIBUTIONS

ST performed the proteomics experiments with nanoparticles, the phagocytosis, rhodamine, NO experiments on the RAW264 cell line, and helped in drafting the manuscript. CAG performed the experiments on primary macrophages, and helped in drafting the manuscript. MCa performed the enzymatic experiments, helped in designing the whole study and in drafting the manuscript, and critically revised the manuscript. HD, JMS and AVD performed and interpreted the mass spectrometry identification in the proteomics experiments, and helped in drafting the manuscript. FP and DH performed the transmission microscopy experiments. AH and NHB performed the zeta potential measurements, the scanning microscopy experiments, and helped in drafting the manuscript. MCh helped in perfoming the experiments on primary macrophages. VCF performed the proteomics experiments on copper ions and the glutathione dosage experiments on RAW264 cells. TR conceived and designed the whole study, took part in the proteomics and targeted experiments, and drafted the manuscript.

## ACKNOWLEDGEMENTS

ST thanks the Université Joseph Fourier for a PhD fellowship. The financial support of the CEA toxicology program (Nanobiomet and Nanostress grants) is also gratefully acknowledged. Finally the support of the Labex SERENADE (11-LABX-0064) is also acknowledged.

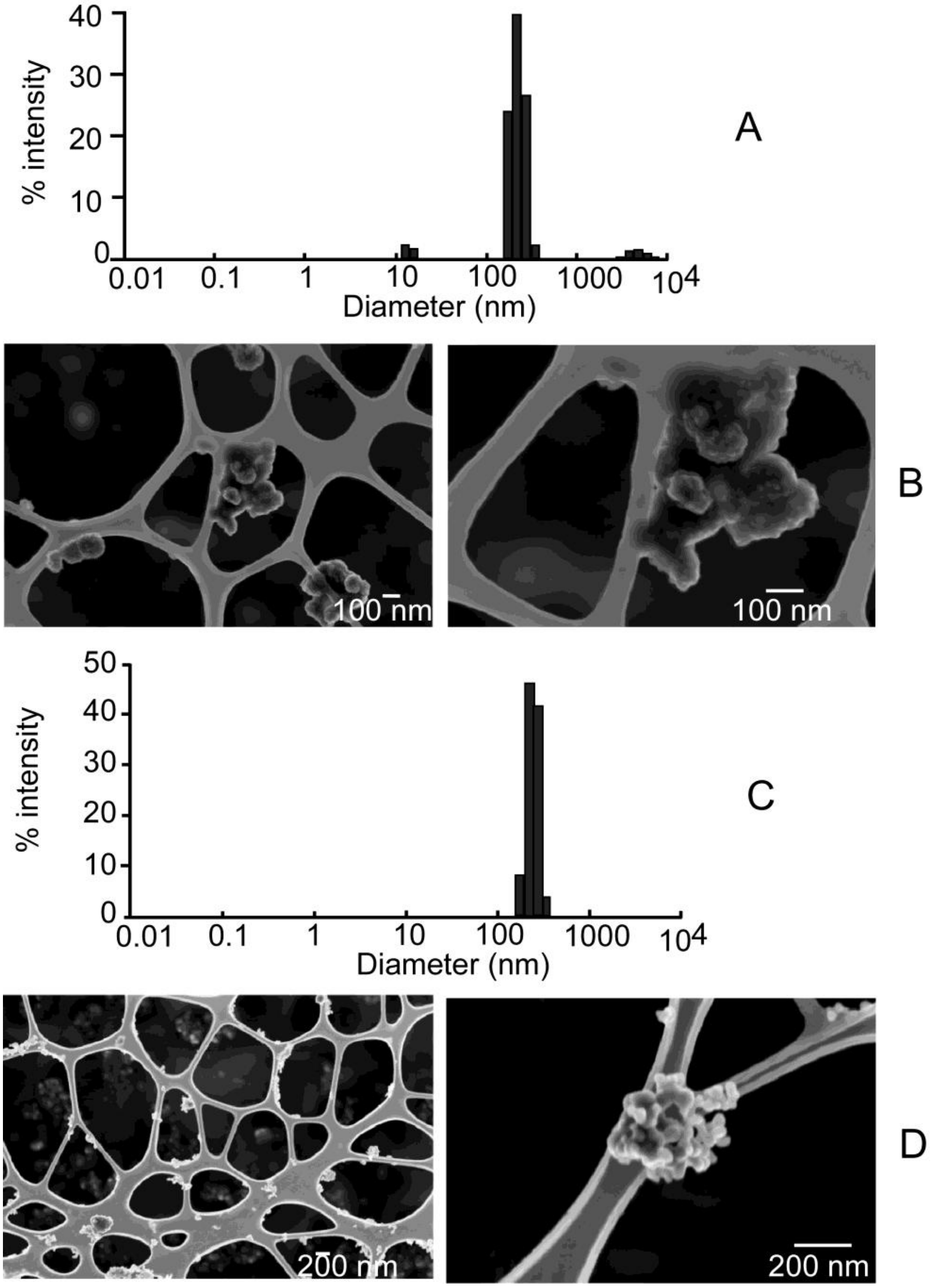

Figure 1: Characterization of the nanoparticles after dispersion by sonication and coating with PVP-40.
A: size distribution of copper nanoparticles after sonication, coating with PVP-40 and dilution in RPMI 1640 medium containing 10% fetal calf serum

B: SEM of copper nanoparticles

C: size distribution of copper oxide nanoparticles after sonication, coating with PVP-40 and dilution in RPMI 1640 medium containing 10% fetal calf serum

D: SEM of copper oxide nanoparticles

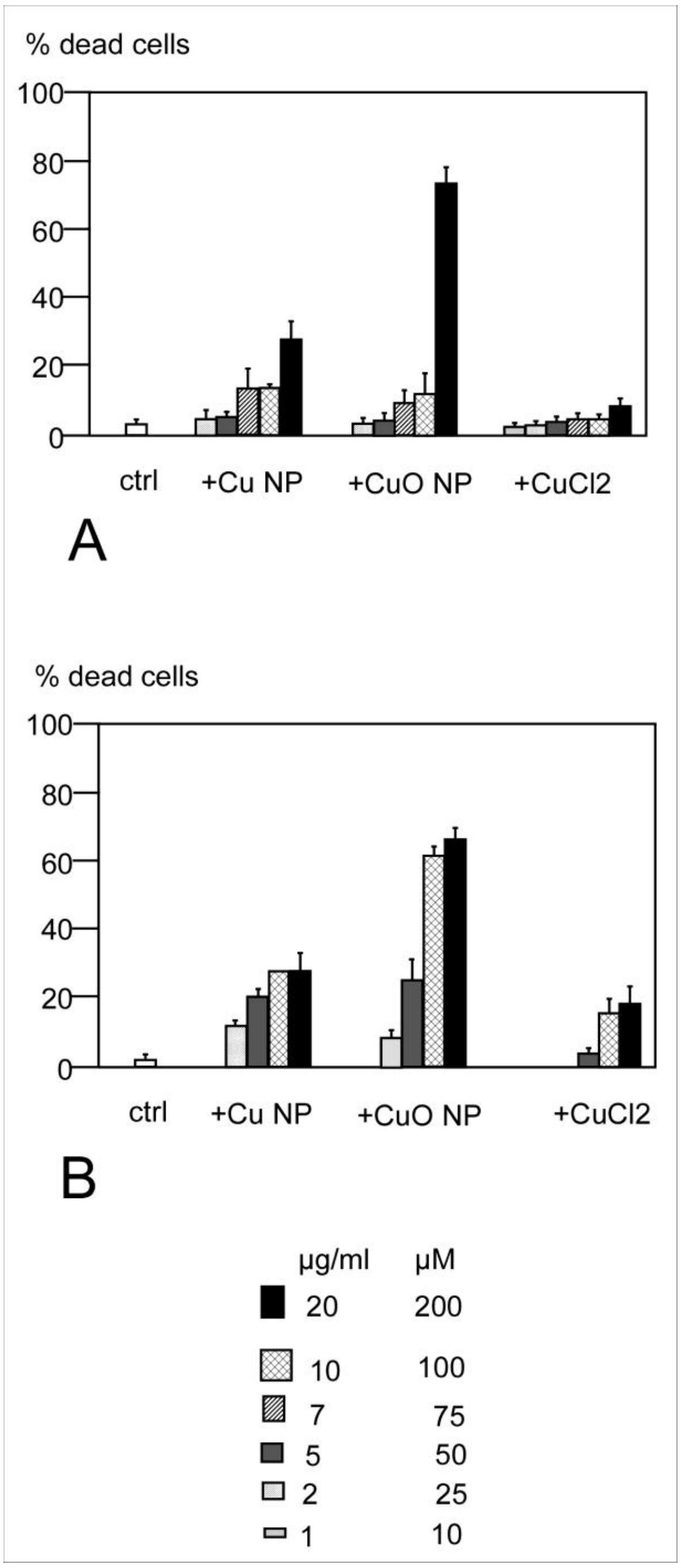

Figure 2: Determination of the viability of the cells after treatment with copper nanoparticles, copper oxide nanoparticles, and copper(II) chloride.

The viability was measured by trypan blue exclusion.

A: RAW 264 cells

B: primary macrophages derived from bone marrow

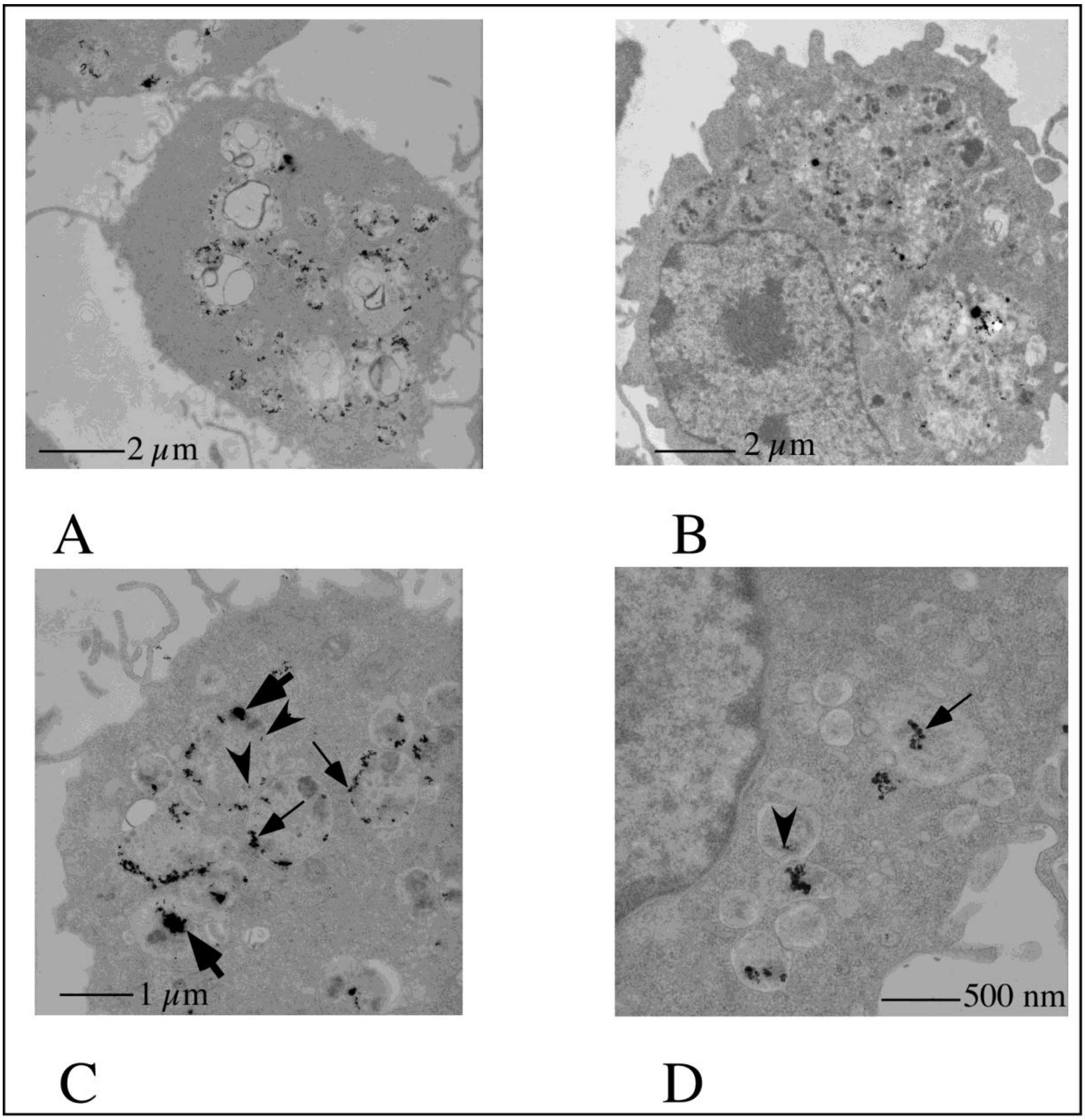

Figure 3: Transmission electron microscopy analysis of nanoparticles-treated cells

RAW264 cells were images by transmission electron microscopy after treatment with copper oxide nanoparticles par 4 hours (panels A, C and D) and 24 hours (panel B). The copper nanoparticles are concentrated in multivesicular bodies and in lysosomes, as expected for professional phagocytic cells. Cells treated for 24 hours show much larger, fusioned lysosomes than cells treated for only 4 hours. Various sizes of copper oxide can be seen, from the 200-300 nm aggregates present in the medium (thick arrows) to the 30-50 nm of the initial (thin arrows) and even to smaller sizes such as 10 nm (arrowheads) as can be expected from the intracellular dissolution of the nanoparticles.

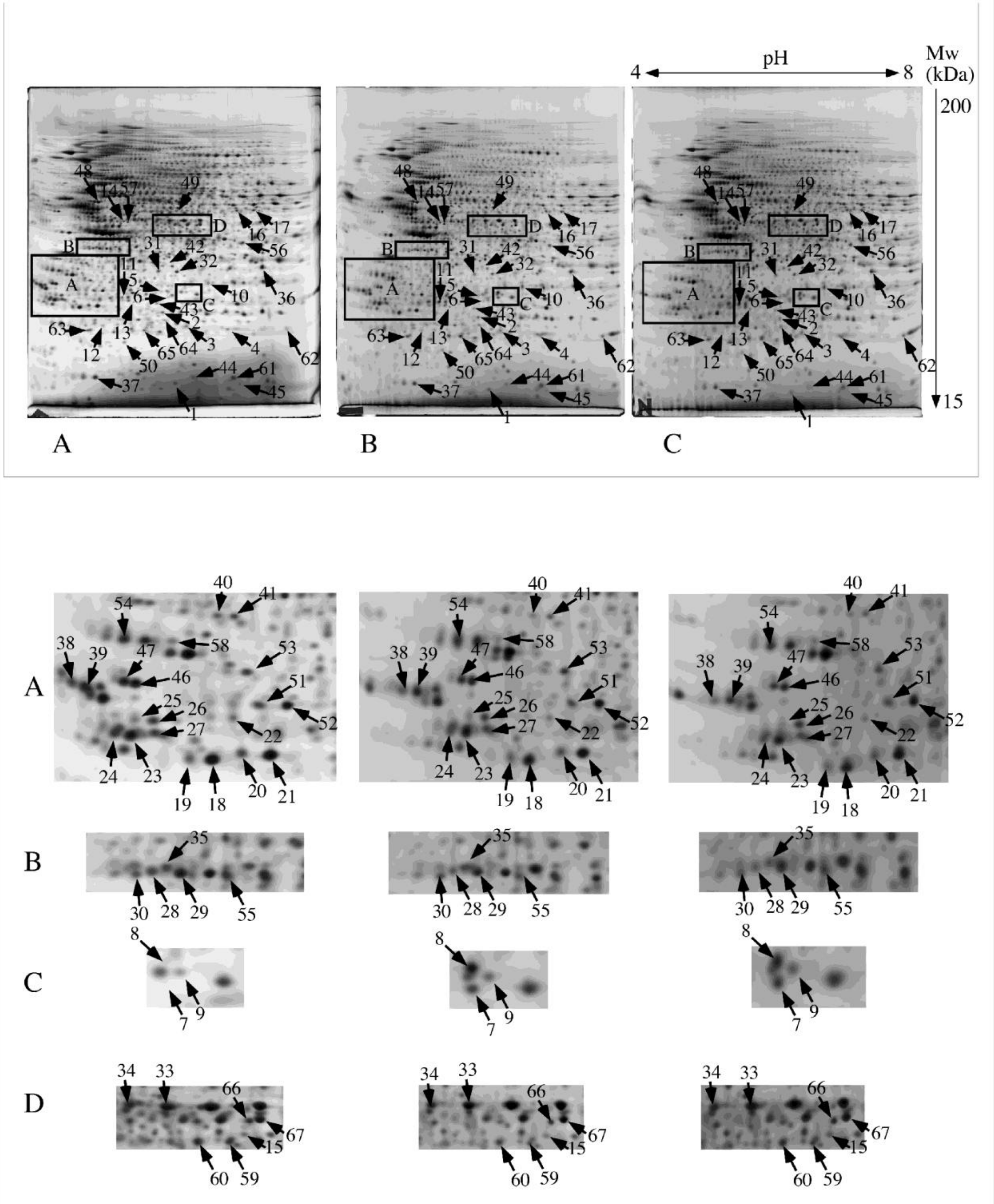

Figure 4: Proteomic analysis by 2D electrophoresis

Part A
Total cell extracts of RAW264 cells were separated by two-dimensional gel electrophoresis. The first dimensions covered a 4-8 pH range and the second dimension a 15-200 kDa range. Total cellular proteins (150 µg) were loaded on the first dimension gel, and the spots were detected by silver staining.
A: gel obtained from control cells
B: gel obtained from copper nanoparticles-treated cells
C: gel obtained from copper oxide nanoparticles-treated cells

The arrows point to spots that show reproducible and statistically significant changes between the control condition and the treatments with Cu or CuO nanoparticles. the numbering of the spots is explained on Table 1.
Rectangles A to D point to areas with a high density of selected spots, shown on a different figure. Gels coming from cells treated with copper chloride have been omitted for clarity reasons

Part B
Insets from gels displayed on Part A, shown at a higher magnification for more clarity in spot pointing

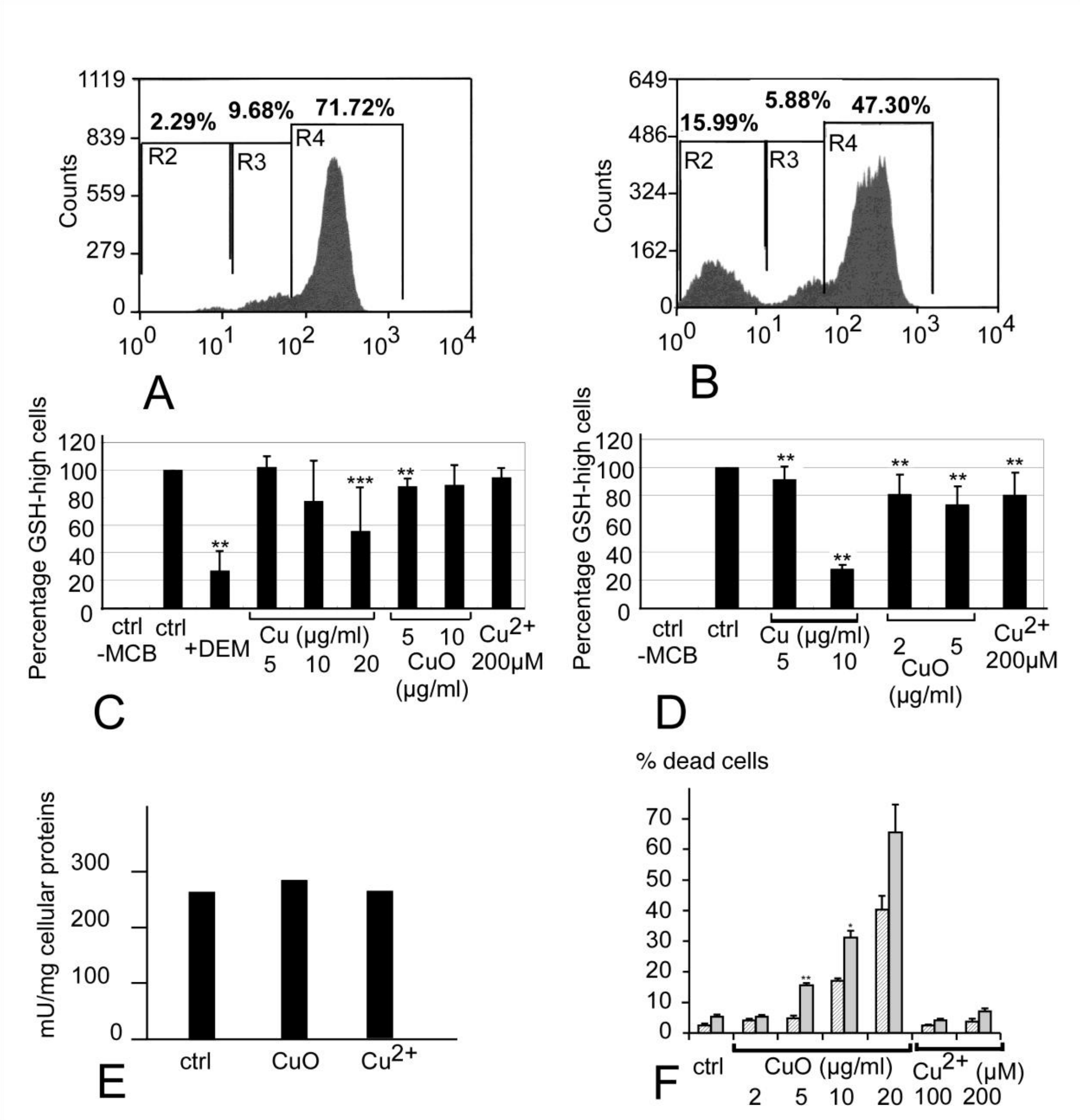

Figure 5: Analysis of the glutathione-based antioxidant system

This figure shows the variation of the intracellular levels of reduced glutathione, estimated through conjugation with monochlorobimane to form a fluorescent adduct and analysis by a MoFlo cell sorter.

Top panel: distribution of the MCB-GSH adduct fluorescence in control cells (A) and cells treated with diethyl maleate (B), which acts as a negative control by reacting with reduced glutathione before the cells are treated with monochlorobimane. This experiment allow to define the GSH-high and GSH-low cells by their fluorescence.

Median panel: variation in the intracellular glutathione levels upon treatment with copper nanoparticles, copper oxide nanoparticles, or copper ion. Three independent experiments per condition

C : variation in the percentage of GSH high cells in RAW264 cells, showing the appearance of a population of GSH-low cells, especially in cells treated with copper nanoparticles.

D: variation in the percentage of GSH high cells in primary macrophages, showing the appearance of a population of GSH-low cells, especially in cells treated with copper nanoparticles.

E: variation in the glutathione reductase activity in RAW264 cells treated with copper oxide nanoparticles or with copper ion

*: $p<0.05$; **: $p<0.01$; ***: $p< 0.001$ (Mann-Whitney test)

F: cell viability upon inhibition of glutathione biosynthesis
Control cells (white bars bars) or cells pre-treated with buthionine sulfoximine to inhibit glutathione biosynthesis grey bars) were treated with various doses of copepr oxide nanoparticles. Viability measured by trypan blue exclusion

*: $p<0.05$; **: $p<0.01$; ***: $p< 0.001$ (Mann-Whitney test)

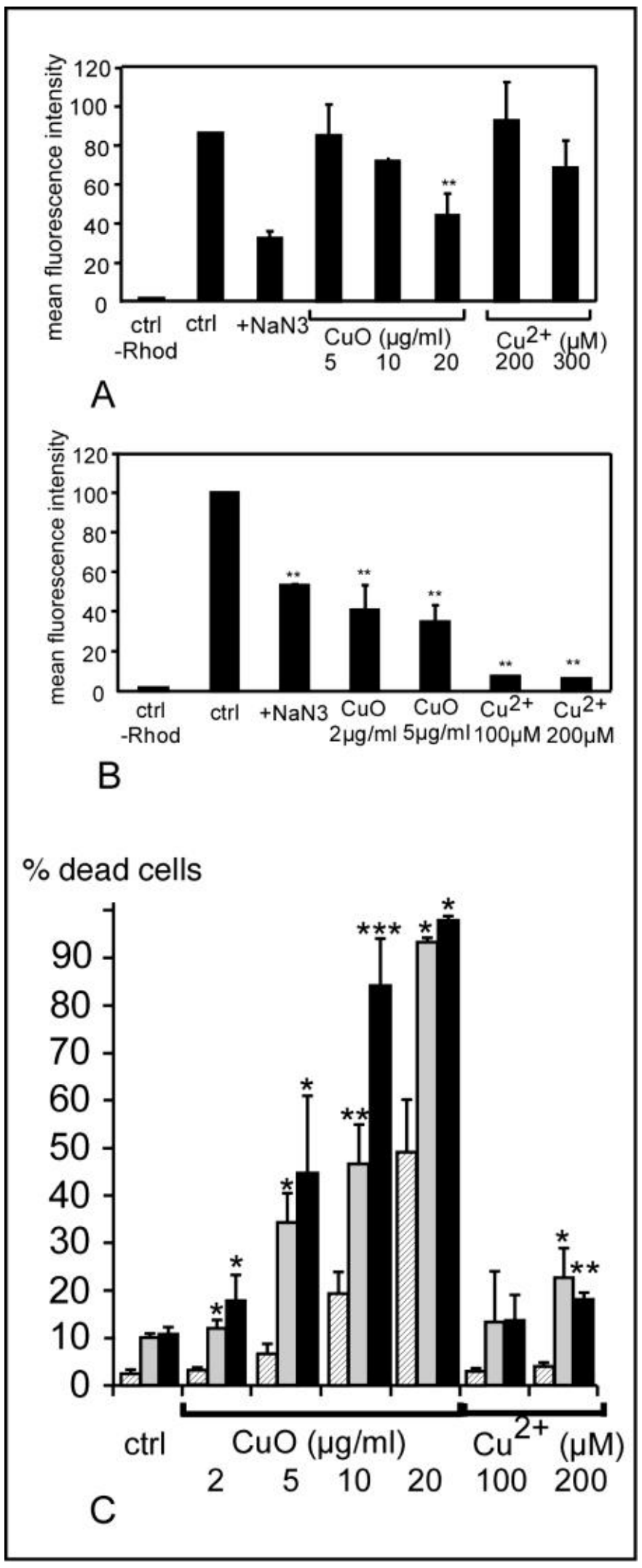

Figure 6: Analysis of the mitochondrial function

The top two panels of this figure show the amount of rhodamine internalized in the cells, expressed in percentage of the fluorescence in control cells (3 independent experiments). Fluorescence measurements were carried out on a FACScalibur cell sorter.
A: results obtained with RAW264 cells
B: results obtained with primary macrophages derived from bone marrow

*: $p<0.05$; **: $p<0.01$ (Mann-Whitney test)

C: cell survival after treatment with inhibitors of the mitochondrial respiratory chain and subsequent treatment with copper oxide nanoparticles.
white bars: control cells; grey bars: cells treated with 200nm rotenone; black bars: cells treated with 200nm antimycin A

*: $p<0.05$; **: $p<0.01$; ***: $p<0.001$ (Student T- test)

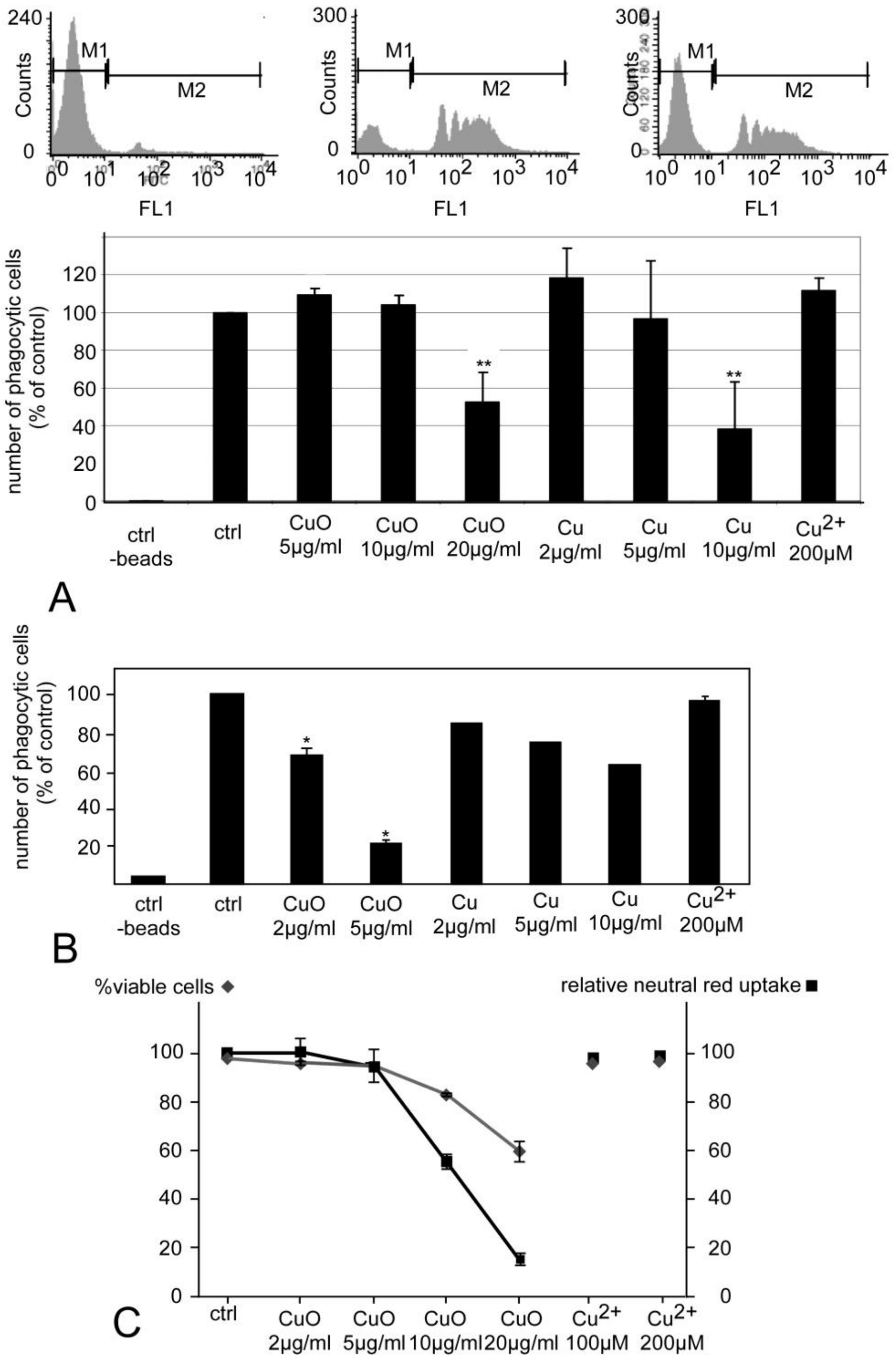

Figure 7: Analysis of the phagocytic capacity

This figure shows the determination of the phagocytic properties of the cells, through ingestion of fluorescent beads and analysis with a FACScalibur cell sorter.

Top panel: distribution of the fluorescence of the cells, showing the negative cells (control at 4°C) and the positive cells (control at 37°C), as well as the raw data obtained for cells treated with copper oxide nanoparticles for 24 hours prior to treatment with the fluorescent beads

Median panel: variation of the phagocytic capacity of the cells under different conditions, expressed in percentage of the control cells. Three independent experiments
A: results obtained with RAW264 cells
B: results obtained with primary macrophages derived from bone marrow (one experiment only for copper nanoparticles)
The results show the strong effect of copper and copper oxide nanoparticles on the phagocytic capacity of the cells, and the absence of effect of the copper ion.

*: $p<0.05$; **: $p<0.01$ (Mann-Whitney test)

Bottom panel
C: comparative variation of cell viability and neutral red uptake capacity.
This figure shows that the lysosomal functionality of the cells is altered by the copper oxide nanoparticles (but not the ion), as the neutral red uptake decreases more steeply than the overall cell viability.

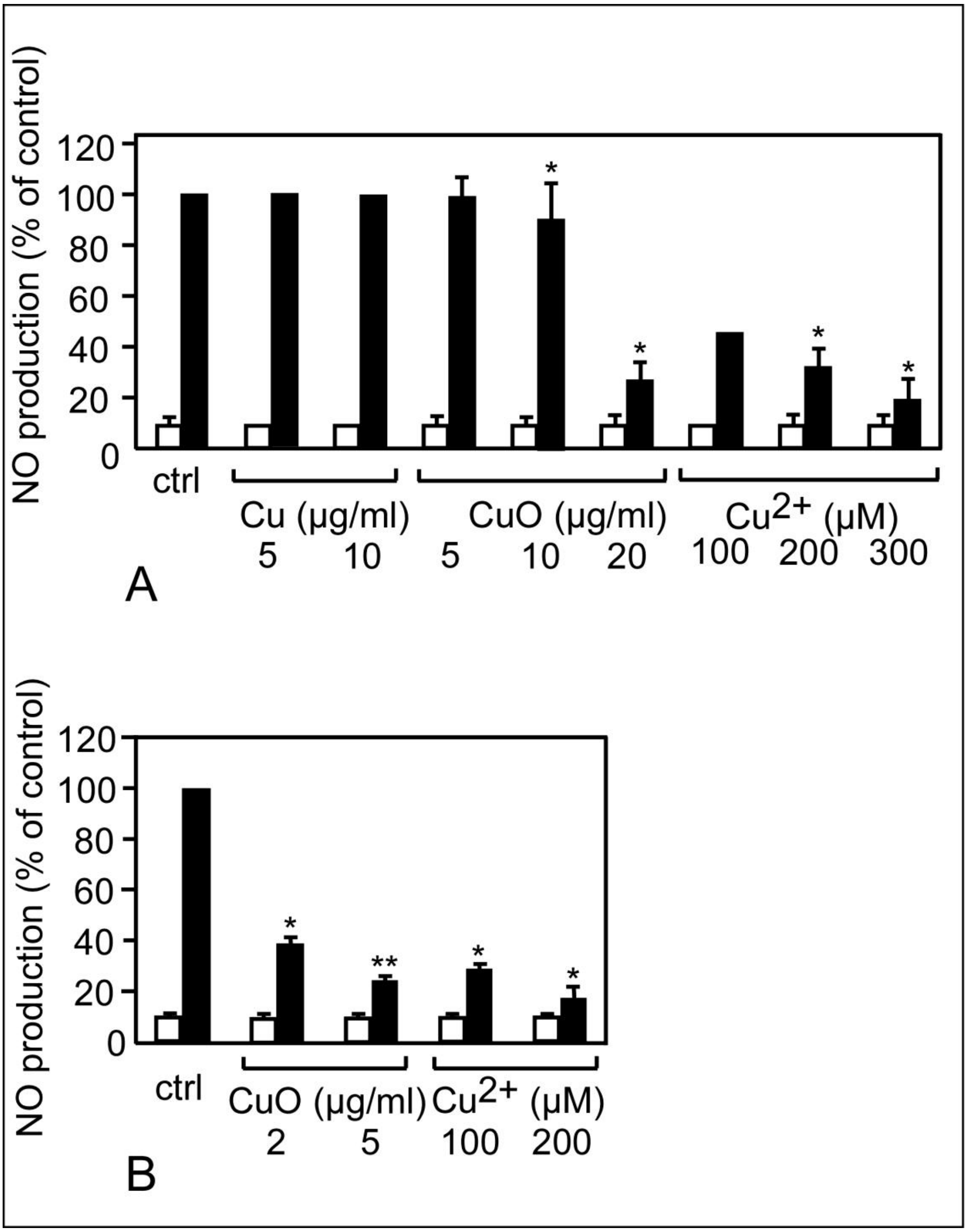

Figure 8: analysis of the nitrous oxide production

This figures shows the amount of NO produced by the cells. White bars: without secondary treatment with LPS. Black bars: with secondary treatment with LPS

A: RAW264 cells
B: Primary macrophages derived from bone marrow
*: $p<0.05$; **: $p<0.01$ (Mann-Whitney test)1.

Table 1: Differentially-expressed proteins identified in the proteomic screen

| Spot number | Protein name | Protein accession numbers | Protein molecular weight (Da) | Cu/ctrl (ratio/p value) | CuO/ctrl (ratio/p value) | Cu2+/ctrl (ratio/pvalue) | ZrO2/ctrl (ratio/p value) | Number of unique peptides |
|---|---|---|---|---|---|---|---|---|
| | **Oxidative stress** | | | | | | | |
| 1 | Superoxide dismutase [Cu-Zn] | P08228 | 15924 | 0.40/0.08 | 0.33 / 0.05 | 1.09/0.72 | 0.824/0.06 | 7 |
| 2 | Peroxiredoxin-6 | O08709 | 24854 | 0.91/0.42 | 0.75/0.04 | 0.91/0.55 | 1.016/0.64 | 5 |
| 3 | Peroxiredoxin-1 | P35700 | 22160 | 2.8/0.04 | 3.3/0.03 | 1.0/1.0 | 1.17/0.21 | 6 |
| 4 | Peroxiredoxin-1 | P35700 | 22160 | 1.28/0.21 | 2.1/0.002 | 1.56/0.01 | 0.921/0.25 | 7 |
| 5 | Heme oxygenase 1 | P14901 | 32911 | 3.8/0.05 | 3.8/0.02 | 0.81/0.52 | 0805/0.82 | 1 |
| 6 | Heme oxygenase 1 | P14901 | 32911 | 1.47/0.12 | 1.61/0.02 | 1.00/0.97 | 0.732/0.25 | 2 |
| 7 | Heme oxygenase 1 | P14901 | 32911 | 3.2/0.05 | 1.49/0.23 | 5.35/0.02 | | 2 |
| 8 | Heme oxygenase 1 | P14901 | 32911 | 1.96/0.001 | 1.93/0.02 | 2.25/0.04 | 1.265/0.05 | 4 |
| 9 | Heme oxygenase 1 | P14901 | 32911 | 1.7/0.22 | 2.1/0.11 | 1.16/0.75 | 1.049/0.70 | 1 |
| 10 | S-formylglutathione hydrolase | Q9R0P3 | 31302 | 1.47/0.03 | 1.84/0.03 | 1.23/0.41 | 0.958/0.70 | 2 |
| 11 | Glutamate--cysteine ligase regulatory subunit | O09172 | 30517 | 1.3/0.04 | 1.35/0.01 | 2.38/0.002 | 0.929/0.15 | 3 |
| | **Mitochondrion** | | | | | | | |
| 12 | NADH dehydrogenase iron-sulfur protein 8, mitochondrial | Q8K3J1 | 24039 | 0.97/0.6 | 0.6/0.002 | 0.98/0.96 | 0.89/0.14 | 8 |
| 13 | NADH dehydrogenase iron-sulfur protein 3, mitochondrial | Q9DCT2 | 30131 | 1.55/0.03 | 1.1/0.5 | 0.8/0.15 | 0.86/0.06 | 4 |
| 14 | Cytochrome b-c1 complex subunit 1, mitochondrial | Q9CZ13 | 52834 | 1.4/0.005 | 1.4/0.007 | 0.72/0.11 | 1.142/0.21 | 6 |
| 15 | Elongation factor Tu, mitochondrial | Q8BFR5 | 49491 | 1.3/0.27 | 1.6/0.05 | 1.16/0.31 | 0.797/0.06 | 3 |
| 16 | Glutamate dehydrogenase 1, mitochondrial | P26443 | 61320 | 1.016/0.92 | 1.345/0.1 | 1.1/0.56 | 0.99/0.88 | 2 |
| 17 | Glutamate dehydrogenase 1, mitochondrial | P26443 | 61320 | 1.32/0.01 | 1.44/0.04 | 0.77/0.19 | 1.095/0.82 | 2 |
| | **Cell signalling** | | | | | | | |
| 18 | Rho GDP-dissociation inhibitor 2 | Q61599 | 22833 | 0.73/0.02 | 0.70/0.01 | 1.048/0.74 | 0.973/0.48 | 3 |
| 19 | Rho GDP-dissociation inhibitor 2 | Q61599 | 22833 | 0.77/0.25 | 0.606/0.07 | 0.94/0.8 | 0.902/0.02 | 3 |
| 20 | Rho GDP-dissociation inhibitor 1 | Q99PT1 | 23390 | 0.67/0.004 | 0.73/0.04 | 0.70/0.09 | 0.947/0.50 | 4 |
| 21 | Rho GDP-dissociation inhibitor 1 | Q99PT1 | 23390 | 081/0.07 | 0.78/0.02 | 0.95/0.54 | 0.977/0.40 | 7 |
| 22 | Inositol monophosphatase 1 | O55023 | 30437 | 0.66/0.003 | 0.67/0.01 | 0.97/0.74 | 1.065/0.46 | 8 |
| 23 | 14-3-3 protein zeta/delta | P63101 | 27772 | 0.67/0.06 | 0.4/0.05 | 0.80/0.25 | 0.933/0.16 | 10 |
| 24 | 14-3-3 protein beta/alpha | Q9CQV8 | 28069 | 0.77/0.11 | 0.66/0.04 | 0.84/0.55 | 0.971/0.80 | 2 |
| 25 | 14-3-3 protein gamma | P61982 | 28285 | 0.805/0.004 | 0.77/0.06 | 0.74/0.60 | 1.088/0.42 | 5 |
| 26 | 14-3-3 protein gamma | P61982 | 28285 | 0.77/0.2 | 0.49/0.015 | 0.88/0.66 | 1.09/0.32 | 7 |
| 27 | 14-3-3 protein beta/alpha | Q9CQV8 | 28069 | 0.89/0.45 | 0.91/0.67 | 0.91/0.68 | 0.889/0.25 | 4 |
| 28 | Guanine nucleotide-binding protein G(i) subunit alpha-2 | P08752 | 40472 | 0.6/0.01 | 0.55/0.005 | 0.89/0.23 | 0.882/0.50 | 4 |
| 29 | Guanine nucleotide-binding protein G(i) subunit alpha-2 | P08752 | 40472 | 0.8/0.06 | 0.78/0.11 | 0.88/0.29 | .0924/0.32 | 5 |
| 30 | Serine-threonine kinase receptor-associated protein | Q9Z1Z2 | 38425 | 0.75/0.02 | 0.65/0.01 | 0.98/0.94 | 0.946/0.52 | 4 |
| | Central metabolism | | | | | | | |
| 31 | Malate dehydrogenase, cytoplasmic | P14152 | 36512 | 0.61/0.045 | 0.62/0.01 | 0.9/0.53 | 0.906/0.19 | 5 |
| 32 | Malate dehydrogenase, | P14152 | 36512 | 0.79/0.2 | 0.63/0.03 | 1.077/0.61 | 0.88/0.11 | 9 |
| | cytoplasmic | | | | | | | |
| 33 | Alpha-enolase | P17182 | 47124 | 0.96/0.65 | 0.82/0.05 | 0.87/0.54 | 0.76/0.08 | 3 |
| 34 | Alpha-enolase | P17182 | 47124 | 0.9/0.3 | 0.8/0.05 | 0.85/0.36 | 0.942/0.64 | 3 |
| 35 | Galactokinase | Q9R0N0 | 42158 | 0.77/0.015 | 0.5/0.001 | 0.91/0.33 | 1.011/0.85 | 5 |
| 36 | L-lactate dehydrogenase A chain | P06151 | 36481 | 0.8/0.03 | 0.85/0.11 | 0.80.01 | 1.066/0.23 | 7 |
| | **Protein production and folding** | | | | | | | |
| 37 | Eukaryotic translation initiation factor 5A-1 | P63242 | 16815 | 0.84/0.2 | 0.43/0.01 | 1.07/0.50 | 0.874/0.14 | 2 |
| 38 | Elongation factor 1-beta | O70251 | 24676 | 0.7/0.06 | 0.5/0.015 | 0.7/0.45 | 1.023/0.87 | 5 |
| 39 | Elongation factor 1-beta | O70251 | 24676 | 1.1/0.57 | 0.8/0.5 | 0.66/0.45 | 1.14/0.23 | 2 |
| 40 | Elongation factor 1-delta | P57776 | 31275 | 0.70/0.07 | 0.48/0.01 | 1.033/0.93 | 1.167/0.21 | 4 |
| 41 | Elongation factor 1-delta | P57776 | 31275 | 0.79/0.16 | 0.7/0.1 | 0.97/0.94 | 1.059/0.27 | 4 |
| 42 | 60S acidic ribosomal protein P0 | P14869 | 34256 | 1.98/0.05 | 2.3/0.001 | 2.6/0.007 | 0.814/0.27 | 2 |
| 43 | Endoplasmic reticulum resident protein 29 | P57759 | 28807 | 1.33/0.08 | 1.78/0.01 | 0.88/0.46 | 0.875/0.80 | 5 |
| 44 | Peptidyl-prolyl cis-trans isomerase A | P17742 | 17954 | 0.66/0.23 | 0.5/0.02 | 1.23/0.57 | 0.828/0.20 | 2 |
| 45 | Peptidyl-prolyl cis-trans isomerase A | P17742 | 17954 | 0.5/0.03 | 0.65/0.15 | 0.88/0.34 | 0.993/0.92 | 1 |
| | **Cytoskeleton and intracellular traffic** | | | | | | | |
| 46 | Tropomyosin alpha-3 chain | Q63610 | 28989 | 0.5/0.07 | 0.45/0.05 | 0.87/0.30 | 1.15/0.24 | 6 |
| 47 | Tropomyosin alpha-3 chain | Q63610 | 28989 | 0.84/0.49 | 0.825/0.45 | 1.03/0.67 | 0.845/0.03 | 12 |
| 48 | Vimentin | P20152 | 53671 | 0.9/0.22 | 0.77/0.05 | 0643/0.13 | 0.944/0.40 | 8 |
| 49 | V-type proton ATPase subunit H | Q8BVE3 | 55838 | 0.77/0.09 | 0.69/0.01 | 1.054/0.80 | 0.835/0.12 | 2 |
| 50 | Transmembrane emp24 domain-containing protein 2 | Q9R0Q3 | 22687 | 1.55/0.04 | 1.6/0.01 | 1.13/0.77 | 1.135/0.26 | 3 |
| | **Miscellaneous** | | | | | | | |
| 51 | Chloride intracellular channel protein 1 | Q9Z1Q5 | 26996 | 0.79/0.09 | 0.71/0.03 | 1.09/0.58 | 0.85/0.14 | 4 |
| 52 | Chloride intracellular channel protein 1 | Q9Z1Q5 | 26996 | 0.86/0.23 | 0.7/0.03 | 0.98/0.57 | 0.982/0.61 | 5 |
| 53 | EF-hand domain-containing protein D2 | Q4FZY0 | 26742 | 0.87/0.21 | 0.69/0.008 | 1.025/0.84 | 1.038/0.62 | 2 |
| 54 | Proliferating cell nuclear antigen | P17918 | 28768 | 0.75/0.06 | 0.66/0.02 | 0.99/0.95 | 0.984/0.85 | 2 |
| 55 | Farnesyl pyrophosphate synthase | Q920E5 | 40565 | 0.8/0.001 | 0.8/0.04 | 0.8/0.14 | 0.984/0.87 | 2 |
| 56 | Poly(rC)-binding protein 1 | O19048 | 37480 | 0.76/0.02 | 0.79/0.003 | 1.2/0.24 | 1.084/0.46 | 1 |
| 58 | Ubiquitin thioesterase OTUB1 | Q7TQI3 | 31253 | 0.77/0.04 | 0.66/0.04 | 1.02/0.83 | 1.015/0.80 | 2 |
| 59 | Adenosylhomocysteinase | P50247 | 47671 | 0.47/0.004 | 0.65/0.02 | 0.9/0.51 | 0.958/0.47 | 11 |
| 60 | Adenosylhomocysteinase | P50247 | 47671 | 0.77/0.01 | 0.78/0.008 | 1.14/0.03 | 1.056/0.65 | 21 |
| | **Other proteins** | | | | | | | |
| 61 | Peptidyl-prolyl cis-trans isomerase A | P17742 | 17954 | 0.75/0.15 | 0.78/0.18 | 1.012/0.95 | 0.981/0.74 | 13 |
| 62 | Superoxide dismutase [Mn], mitochondrial | P09671 | 24585 | 1.17/0.22 | 1.08/0.75 | 0.38/0.11 | 1.347/0.53 | 4 |
| 63 | Peroxiredoxin-2 | Q61171 | 21761 | 0.92/0.62 | 0.70/0.23 | 1.03/0.72 | 1.02/0.73 | 7 |
| 64 | Peroxiredoxin-3 | P20108 | 28109 | 1.41/0.10 | 1.30/0.22 | 1.13/0.12 | 0.976/0.73 | 6 |
| 65 | Ferritin light chain 1 | P29391 | 20785 | 1.29/0.09 | 1.31/0.25 | 1.62/0.025 | 1.18/0.05 | 9 |
| 66 | Eukaryotic initiation factor 4A-III | P38919 | 46824 | 1.046/0.75 | 1.13/0.38 | 0.97/0.88 | 0.933/0.55 | 20 |
| 67 | Elongation factor 1-gamma | Q9D8N0 | 50043 | 1.004/0.98 | 0.952/0.73 | 1.0/0.99 | 0.964/0.65 | 16 |

*The accession numbers are those of the SwissProt Database.